\documentstyle[aps,prl,epsfig,multicol,times,amsfonts]{revtex}
\makeatletter
\newlength\abovecaptionskip \newlength\belowcaptionskip
\def\@makecaption#1#2{%
 \vskip\abovecaptionskip \sbox\@tempboxa{#1: #2}%
 \ifdim \wd\@tempboxa >\hsize #1: #2\par \else \global \@minipagefalse
 \hb@xt@\hsize{\hfil\box\@tempboxa\hfil}%
 \fi \vskip\belowcaptionskip} \makeatother

\newcommand{\lan}{\langle}
\newcommand{\ran}{\rangle}

\newcommand{\eps}{\varepsilon}
\newcommand{\beq}{\begin{equation}}
\newcommand{\eeq}{\end{equation}}
\newcommand{\dav}{\left<d\right>}

\title{Periodic orbit theory and spectral rigidity in pseudointegrable systems}
\author{Jesper Mellenthin and Stefanie Russ\\
Institut f\"ur Theoretische Physik III, Universit\"at Giessen, \\
D-35392 Giessen, Germany}
\date{\today}
\begin{document}            

\draft
\maketitle
\begin{multicols}{2}[%
\begin{abstract} 
We calculate numerically the periodic orbits of pseudointegrable systems of low genus numbers $g$ that arise from rectangular systems with one or two salient corners. From the periodic orbits, we calculate the spectral rigidity $\Delta_3(L)$ using semiclassical quantum mechanics with $L$ reaching up to quite large values. 
We find that the diagonal approximation is applicable when averaging over a suitable energy interval. 
Comparing systems of various shapes we find that our results agree well with
$\Delta_3$ calculated directly from the eigenvalues by spectral statistics. 
Therefore, additional terms as e.g. diffraction terms seem to be small in the case of the systems investigated in this work.
By reducing the size of the corners, the spectral statistics of our pseudointegrable systems approaches the one of an integrable system, whereas very large differences between integrable and pseudointegrable systems occur, when the salient corners are large. 
Both types of behavior can be well understood by the properties of the periodic orbits in the system.
\end{abstract}
]

\section{Introduction}

The motion of a classical particle in a billiard system can show regular, 
chaotic or intermediate behavior, depending on the billiard geometry.
In a chaotic billiard, the motion is ergodically extended over the whole energy surface in phase space and two particles whose trajectories are very close at the beginning, diverge exponentially from each other. 
If the system is integrable on the other hand, the motion of the billiard particle is restricted to a two-dimensional torus in phase space and neighboring trajectories diverge only linearly from each other. Examples for chaotic billiards are e.g. the Sinai or the stadium billiard, whereas rectangular or circular billiards are integrable.
Between these two limiting cases, there are several classes of intermediate systems.
 
A potential well of the same geometry as the corresponding classical billiard 
-- a quantum billiard -- reflects this regular, chaotic or intermediate 
behavior in the statistics and the dynamics of its eigenvalues.
The statistics of the eigenvalues investigates the static correlations of 
the eigenvalues and the distribution of the distances between consecutive 
values \cite{mehtabuch}. It can be determined directly by calculating first the
eigenvalues and then their distribution and the correlations between them. 
It can also be determined at least approximately by calculating the periodic orbits of the system and applying semiclassical quantum mechanics. 

In this paper, we want to use the periodic orbit theory.
We focus on pseudointegrable systems \cite{pseudo1,pseudo2,pseudo3}, which are an interesting example of an intermediate class. 
Like in integrable systems, the motion of a classical particle in a 
pseudointegrable system is restricted to a two-dimensional surface in phase 
space. However, these surfaces do not have the shapes of tori but are 
more complicated objects with more than one hole (''multihandeled spheres'').
Examples for pseudointegrable systems are polygons with only rational angles 
$n_i\pi/m_i$, with $n_i, m_i\in \mathbb{N}$ and at least one $n_i>1$. 
They are classified by their genus number 
\begin{equation}\label{genus}
g = 1 + \frac{M}{2} \sum_{i=1}^{J}\frac{n_i-1}{m_i},
\end{equation}
which is equal to the number of holes in the multihandeled sphere in
phase space. Here, $J$ is the number of angles 
and $M$ is the least common multiple of the $m_i$.
The reason why those systems are not completely integrable is their property of 
beam splitting. At some points in their geometry, neighboring trajectories of 
particles can be split into two opposite directions. Fig.~\ref{bi:geo} shows
examples of pseudointegrable billiards with genus numbers $g=2$ and $3$. 
The beam splitting property is demonstrated in Fig.~\ref{bi:geo}(a) 
at one of the salient corners.

The first possibility to calculate the spectral statistics starts with the eigenvalues. Here, one first looks at the distribution $P(s)$ of the normalized distances $s_i=(\eps_{i+1}-\eps_i)/\lan s \ran$, $\lan s \ran=1$, between two consecutive energy levels $\eps_{i+1}$ and $\eps_i$ with the mean distance $\lan s \ran$, which has two limiting cases. When the systems are integrable, $P(s)$ follows the Poisson distribution, whereas the $s_i$ of chaotic systems are Wigner-distributed.
The distribution $P(s)$ of pseudointegrable systems has been found to be intermediate between both \cite{cc,shudoshim,shudoetal,steffi,ys2003,syprog} and it is assumed that with increasing $g$ they come closer to the behavior of chaotic systems. However, it is not yet clear, in which way other system details interfere.

As a measure for the correlations between the eigenvalues, we consider the 
spectral rigidity $\Delta_3(L)$ \cite{dyson}, which describes the correlations in a normalized energy interval of length $L$. $L$ gives the approximate number of energy levels in the considered interval and will in the following be called the argument of $\Delta_3$. We start from the integrated density of states
$N(\eps)=\sum_{n=1}^N \Theta(\eps-\eps_n)$ of the normalized (''unfolded'') energy levels, which is a staircase and can be approximated by a straight line. 
$\Delta_3(L)$ is defined as the least square deviation,
\begin{equation}\label{delta3}  
\Delta_3(L) = \left\lan\rm{Min}_{r_1,r_2} \int_{E_0-L/2}^{E_0+L/2} 
[N(\eps)-r_1-r_2\eps]^2 d\eps  
\right\ran,  
\end{equation}  
where $\rm{Min}_{r_1,r_2}$ means that the parameters $r_1$ and $r_2$ are 
chosen such that the line $r_1+r_2\eps$ is the best fit of $N(\eps)$. The average $\lan\dots\ran$ is an energy average, carried out over many different values of $E_0$ in an interval $\Delta E$. $\Delta E$ should not be confused with the argument $L$, which gives the length of the considered energy interval.
The limiting curves for not too large $L$ are $\Delta_3(L)=L/15$ for integrable systems and $\Delta_3(L)=\ln(L)/\pi^2-0.07/\pi^2 + O(L^{-1})$ for the ensemble of Gaussian orthogonal matrices (GOE) \cite{mehtabuch,berry1}, which serves as a 
generally accepted good limit for chaotic systems. 
For pseudointegrable billiards, $\Delta_3(L)$ has been found intermediate between both (see above).

\unitlength 1.85mm
\vspace*{0mm}
\begin{figure}
\begin{picture}(0,30)(0,10)
\def\epsfsize#1#2{0.5#1}

\put(0,38){\makebox(1,1){(a)}}
\put(1,11){\makebox{\includegraphics[width=3.3cm, height=5.2cm]{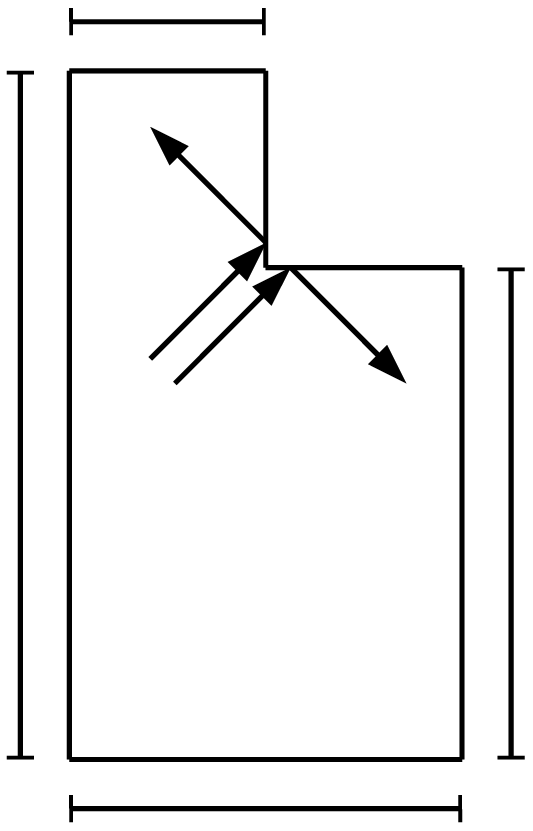}}}

\put(9.7,10.7){\makebox(1,1){$X_1$}}
\put(6.2,38.9){\makebox(1,1){$X_2$}}
\put(0,25){\makebox(1,1){$Y_1$}}
\put(19,22){\makebox(1,1){$Y_2$}}

\put(21,38){\makebox(1,1){(b)}}
\put(23,11.5){\makebox{\includegraphics[width=3.3cm, height=5.08cm]{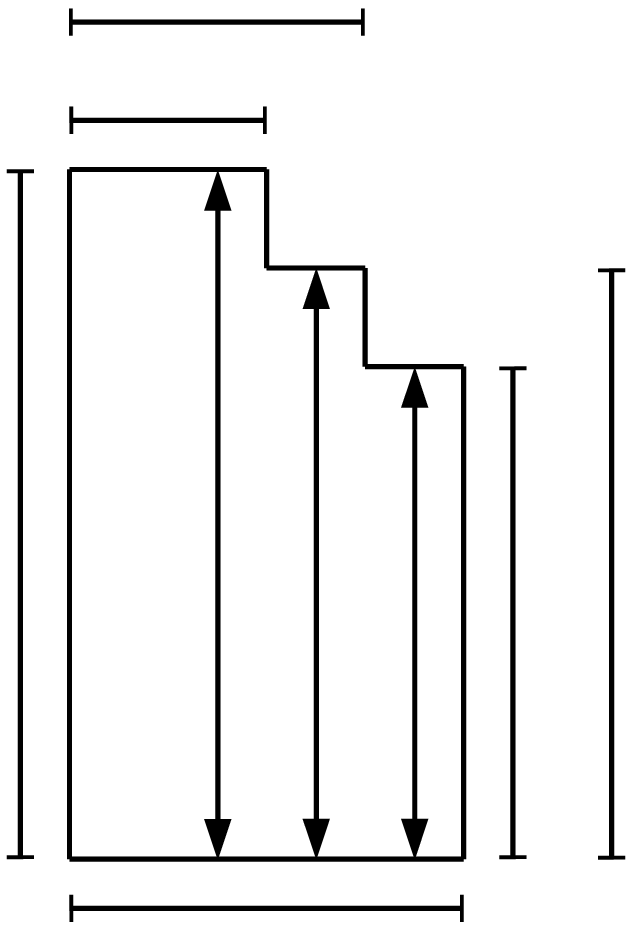}}}

\put(30.5,10.7){\makebox(1,1){$X_1$}}
\put(28.3,38.8){\makebox(1,1){$X_2$}}
\put(27.3,36){\makebox(1,1){$X_3$}}
\put(22.1,24){\makebox(1,1){$Y_1$}}
\put(41,22){\makebox(1,1){$Y_2$}}
\put(38.3,20.8){\makebox(1,1){$Y_3$}}

\end{picture}
\caption[]{\small Shapes of the pseudointegrable systems 
considered in this paper: (a) the one-step system $(X_1,Y_1;X_2,Y_2)$ 
with genus number $g=2$ and (b) the two-step system $(X_1,Y_1;X_2,Y_2;X_3,Y_3)$
with $g=3$. In (a), the beam splitting of two initially neighboring trajectories 
at a salient corner is demonstrated and in (b) trajectories of three 
''neutral'' orbits  (between two parallel walls) are indicated by arrows. For the lengths $X_i$, $Y_i$ considered in this paper see Tab.~\ref{tab:1}.}
\label{bi:geo}
\end{figure}

A second possibility to calculate $\Delta_3(L)$ is given by the periodic orbit theory.
In pseudointegrable systems, all periodic orbits form families of equal lengths and the starting point of an orbit can always be shifted to at least one direction along the boundary without leaving the family. The simplest and shortest orbit families are the ''neutral orbits'' \cite{sieber93} that bounce between two parallel walls.
Using semiclassical quantum mechanics, setting $\hbar=2m=1$ and neglecting additional contributions coming e.g. from diffractive orbits, $\Delta_3(L)$ under Neumann boundary conditions is given by \cite{berry1,biswas1,biswas2,footnote1}, 
\begin{equation}
\label{biswaseq}
\Delta_3(L) =  \left<\frac {\sqrt{E_0}}{4 \pi^3} \sum_{i,j} \frac
{a_ia_j}{\left(\ell_i\ell_j\right)^{3/2}} \cos\left[\sqrt{E_0}(\ell_i-\ell_j)\right] H_{ij}\right>.
\end{equation}
The double sum is carried out over all orbit families 
(including repetitions, but only in forward direction) of lengths $\ell_i$ and $\ell_j$, $a_i$ and $a_j$ are the areas in phase space that are occupied by the respecting orbit families and the function $H_{ij} = F(y_i-y_j)-F(y_i)F(y_j)-3F'(y_i)F'(y_j)$, where $F(y)=(\sin y)/y$ and primes denote differentiation. The argument $L$ enters via $y_i=(L\ell_i)/(4\sqrt{E_0}\left< d\right>)$ with the mean level spacing $1/\dav=4\pi/A$ and the system area $A$. As in Eq.~(\ref{delta3}), the average $\lan\dots\ran$ in Eq.~(\ref{biswaseq}) is carried out over different energies $E_0$ in an energy interval of width $\Delta E$. 

In this paper, we basically want to use Eq.~(\ref{biswaseq}) to calculate $\Delta_3(L)$ and see in which way the different system details apart from the genus number $g$ influence its behavior.
It can be seen by Eq.~(\ref{biswaseq}) that the number of orbits in
the different length intervals $\left[\ell, \ell+\Delta\ell\right]$ as well as the corresponding areas $a(\ell)$ are the important quantities to investigate this question. We therefore carefully calculate these quantities for systems with different lengths and widths $(X_i,Y_i)$ of the different segments (cf. Fig.~\ref{bi:geo}) and see how the behavior of $\Delta_3$ changes by varying the systems. 

The behavior for large $L$ is determined by the short orbits and we will sometimes use the neutral orbits for crude approximations.
By choosing the segments $X_2\to X_1$, $Y_2\to Y_2$, the orbits approach the ones of a rectangular system, which is integrable. For large differences between $X_1$ and $X_2$ and between $Y_1$ and $Y_2$, on the other hand, the orbit families can become very different from those of the rectangle. 

The paper is organized as follows: In section~\ref{sec_Calc_orbits}, we explain how the lengths and areas of the periodic orbits are calculated and show the results. In section~\ref{sec_Diag_approx},
we discuss the applicability of the diagonal approximation, where only terms of $\ell_i=\ell_j$ are considered in Eq.~(\ref{biswaseq}). In section~\ref{sec_Delta3} we finally show the periodic orbits results for $\Delta_3(L)$ for many different systems and compare them to the eigenvalue statistics. Some of the considered systems are very close to integrability, while other systems possess very pronounced salient corners and we show how $\Delta_3(L)$ is influenced by these system details. Finally, in the conclusion section~\ref{sec_Concl}, the influence of neglected terms as e.g. the diffraction terms and possible deviations between the eigenvalue statistics and the periodic orbits results are discussed.

\section{Calculation of the periodic orbits}
\label{sec_Calc_orbits}
It has been shown in \cite{biswas1,biswas2} that for large $\ell$, the 
proliferation rate $N(\ell)$, i.e. the number of orbits with lengths smaller than $\ell$ grows quadratically
with $\ell$, 
\beq\label{proliferat}
N(\ell) = \pi b_0\ell^2/\lan a(\ell)\ran,
\eeq 
where $\lan a(\ell)\ran=\sum_{i,\ell_i<\ell} a_i/\sum_{i,\ell_i<\ell} 1$ is the average
area in phase space, occupied by the orbits with lengths smaller than $\ell$ and $b_0$ is a constant, depending slightly on the details of the system,
\beq 
\label{proliferat_bo}
b_0\approx(1/2\pi)\sum_i^\infty \delta(\ell-\ell_i)\, a_i/\ell_i
\eeq 
(the sum going over all
periodic orbits including repetitions). 
It has been found that $b_0=1/4$ for integrable systems and slightly
larger for pseudointegrable systems, whereas $\lan a(\ell)\ran\approx 4A$ in integrable systems and considerably smaller for pseudointegrable systems \cite{biswas2}. 

\unitlength 1.85mm
\vspace*{0mm}
\begin{figure}
\begin{picture}(0,30)
\def\epsfsize#1#2{0.65#1}
\put(0,0){\epsfbox{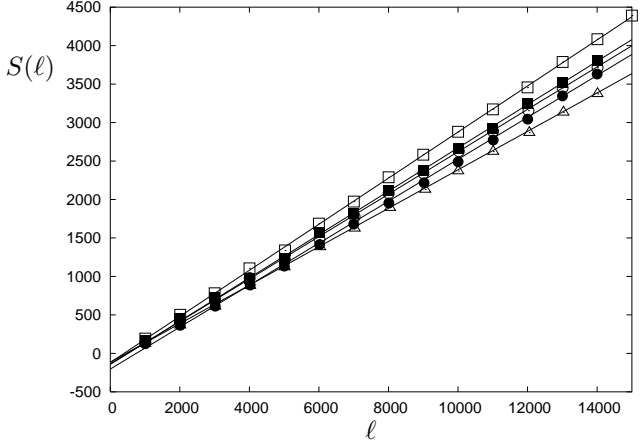}}
\put(23.5,-1.0){\makebox(1,1){$\ell$}}
\put(-1.0,25){\makebox(1,1){$S(\ell)$}}
\end{picture}
\caption[]{\small Test of the sum rule (see Eq.~\ref{sumrule}) for the geometries of Fig.~\ref{bi:geo} with different values of $X_i,Y_j$. $S(\ell)$ is plotted versus $\ell$ (giving straight lines) and $b_0$ is calculated from the slopes. The different symbols refer to the rectangular system (triangular symbols) and to pseudointegrable systems of $g=2$ (circles) and $g=3$ (squares). Open and closed symbols refer to different step sizes (for details of the systems see Tab.~\ref{tab:1}).
For the rectangle, we find $b_0=1/4$ and for the pseudointegrable systems, $b_0$ is slightly increasing with $g$ and with the step sizes. 
}
\label{bi:sumrule}
\end{figure}

We have calculated the periodic orbits of our pseudointegrable systems by two
different methods. 
The first method uses the fact that one member of each orbit family must either start at a salient corner or pass close to it. 
Therefore, we start at the salient corners and vary the reflection angle $\varphi$ between the trajectory and the billiard wall in small 
steps $\Delta\varphi$ until the orbit nearly closes. Finally we 
use an iteration method \cite{DaSilva} to find the exact reflection angle $\varphi$. 
The second method calculates $\varphi$ and the lengths $\ell_\varphi$ of all hypothetical orbits by \cite{biswas3} $\tan \varphi=\sum_i n_i Y_i/\sum_j m_j X_j$, $\ell_\varphi=2 \left[(\sum_i n_i Y_i)^2+(\sum_j m_j X_j)^2\right]^{1/2}$, where $n_i$ and $m_j$ are positive integers and $X_i$, $Y_j$ are the segment lengths as shown in Fig.~\ref{bi:geo}. 
In pseudointegrable systems, due to the shielding role of the corners, not all hypothetical orbits really occur in each system. Therefore, we have to check 
which trajectories actually return to their starting point within the correct length $\ell_\varphi$ of the trajectory. 
Both methods basically lead to the same results. The agreement, e.g. for the one-step system is larger then $97\%$ for the orbits, being reflected at the boundaries up to 50 times. Going to more reflections or to systems with more steps, the iteration method misses more orbits, depending on the value of
$\Delta\varphi$ and the numerical tolerances, which makes the second method more reliable and therefore better suited for this work. 
Moreover, the second method is faster for our systems of small genus numbers, but we assume that the iteration procedure could be favorable in more complicated systems, where the number of hypothetical orbits can become very large. 

Naturally, these numerical procedures are restricted to a maximum orbit length, in
our case to the first $3000$ orbits (for tests $12000$), including repetitions. This is legitimate as
the contributions of the diagonal elements in Eq.~(\ref{biswaseq}) decay with
$\ell^3$ (the contributions of non-diagonal elements can be neglected as
will be shown in the next section) whereas the number of orbits per length interval
$\Delta\ell$ increases only with $2\ell\Delta\ell$ (see Eq.~(\ref{proliferat})). With $H_{ij}$ staying finite, the alternating sum of Eq.~(\ref{biswaseq}) is convergent. 

\unitlength 1.85mm
\vspace*{0mm}
\begin{figure}
\begin{picture}(0,30)
\def\epsfsize#1#2{0.65#1}
\put(0,0){\epsfbox{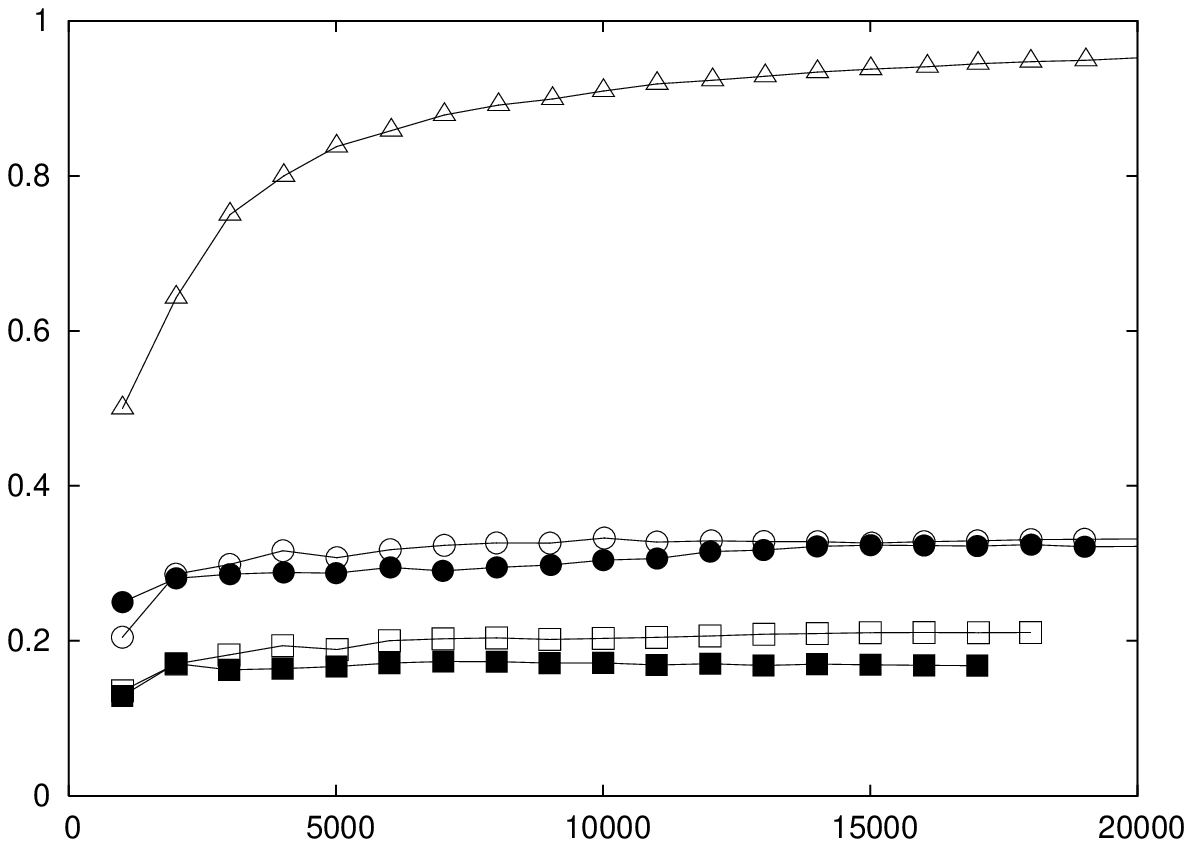}}
\put(23.5,-1.0){\makebox(1,1){$\ell$}}
\put(-0.5,25){\makebox(1,1){$\frac {\lan  a(\ell)\ran}{4A}$}}
\end{picture}
\caption[]{\small Normalized average area $\lan a(\ell)\ran$ in phase space for 
the periodic orbits with lengths smaller than $\ell$ for the same systems (and same symbols) as in
Fig.~\ref{bi:sumrule}. It can be seen that systems of the same genus number $g$ form groups of similar $\lan a(\ell)\ran$. For large values of $\ell$, the areas are saturating.
}
\label{bi:flaeche}
\end{figure}

First, we verify that our periodic orbits fulfill Eq.~(\ref{proliferat}) and  investigate how $b_0$ depends on the system details. $b_0$ can approximately be calculated from the sum rule \cite{biswas2}
\beq\label{sumrule}
S(\ell)\equiv S(\ell_{i_{\rm{max}}}) = \sum_i^{i_{\rm{max}}} a_i/\ell_i\approx 2\pi b_0\ell,
\eeq 
which can be easily verified by replacing the sum by an integral over $d\ell$, inserting the density $dN(\ell)/d\ell$ with $N(\ell)$ from Eq.~(\ref{proliferat}) and replacing the areas $a_i$ by their mean value $\lan a\ran$. 
In Fig.~\ref{bi:sumrule}, we plot $S(\ell)$ versus $\ell$ for the
pseudointegrable systems of Fig.~\ref{bi:geo} with different step sizes as 
well as for the rectangular system. In all cases, $S(\ell)$
increases linearly with the upper orbit length $\ell$ and $b_0$ is determined from the slopes by a least square fit. We can see that $b_0$ increases slightly with the genus number $g$ and the step sizes.

Next, we calculate the average area $\lan a(\ell)\ran$ in phase space for 
the periodic orbits with lengths smaller than $\ell$. In Fig.~\ref{bi:flaeche}, we
plot the normalized average area $\lan a(\ell)\ran/(4A)$ versus $\ell$. In 
agreement with previous calculations \cite{biswas2,biswas3}, the
values of $\lan a(\ell)\ran$ are saturating for large values of $\ell$
and are considerably smaller than the value of 
$\lan a(\ell)\ran\approx 4A$ of integrable systems. 
Moreover, it can be seen that systems of the same genus number $g$
form groups of similar $\lan a(\ell)\ran$. This can be most easily understood by
comparing the neutral orbits of Fig.~\ref{bi:geo}(b): If we start with a rectangle and disturb it by one salient corner, each neutral orbit will split into two different families, each of them covering a smaller area than before. The same effect also occurs for more complex orbits and is repeated with each additional salient corner. So, the average area $\lan a(\ell)\ran$ will be reduced with $g$, whereas the number of different orbit families increases.

Now, we can calculate the proliferation rate $N(\ell)$ for our systems according to Eq.~(\ref{proliferat}) by using the values of $b_0$ and 
$\lan a(\ell)\ran$ from Figs.~\ref{bi:sumrule} and \ref{bi:flaeche}, respectively.
In Fig.~\ref{bi:proliferat} we compare the values from Eq.~(\ref{proliferat}) 
(straight lines) to the numbers of $N(\ell)$, obtained by counting the different orbits (symbols). The agreement between both curves is excellent. Clearly, $N(\ell)$ increases much faster for higher values of $g$ than for lower values (see above).
Therefore, also $N(\ell)$ forms groups of very close-lying curves that correspond to systems with the same $g$. Like $\lan a(\ell)\ran$ also $N(\ell)$ is basically determined by $g$ and changes only very slightly by further details of the considered systems.

\unitlength 1.85mm
\vspace*{0mm}
\begin{figure}
\begin{picture}(0,30)
\def\epsfsize#1#2{0.65#1}
\put(0,0){\epsfbox{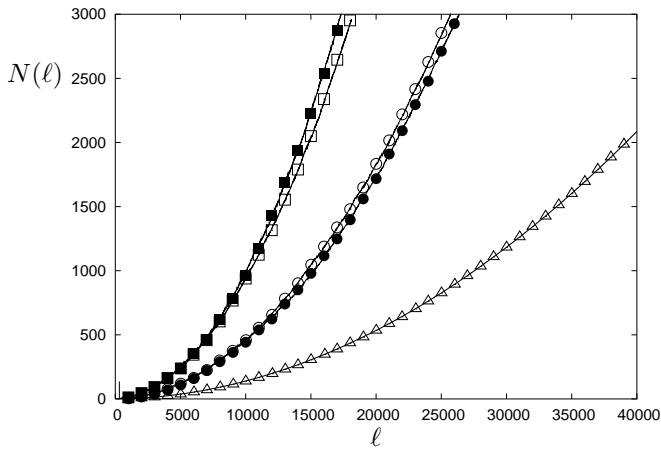}}
\put(23.5,-1.0){\makebox(1,1){$\ell$}}
\put(-1.0,25){\makebox(1,1){$N(\ell)$}}
\end{picture}
\caption[]{\small The proliferation rate $N(\ell)$ is plotted versus $\ell$ for five different systems and two types of calculations. The lines are calculated 
according to Eq.~(\ref{proliferat}), using the values of $b_0$ and 
$\lan a(\ell)\ran$ from Figs.~\ref{bi:sumrule} and \ref{bi:flaeche}, respectively. The symbols (same symbols as in Fig.~\ref{bi:sumrule}) show the numerical data, gained by a simple summation of the different periodic orbits (including repetitions). Both curves agree very well.
}
\label{bi:proliferat}
\end{figure}

\section{The diagonal approximation}
\label{sec_Diag_approx}
As Eq.~(\ref{biswaseq}) involves a double sum over $\ell_i$ and $\ell_j$,
it is very convenient to use the diagonal approximation, where only terms with
$\ell_i=\ell_j$ are taken into account. 
It has been shown in \cite{berry1} that this is justified for integrable
systems. The reason is that only orbit pairs where $\ell_i-\ell_j\stackrel{<}{\sim} 1/\sqrt{\Delta E}$ survive the average over an energy interval $\Delta E$, while for larger phase differences, the different terms cancel by the oscillations of the cosine function. 
With $N(\ell)\sim\ell^2$ and for not too small $\Delta E$, the number of these pairs grows 
more slowly than their contributions decrease and non-diagonal terms can therefore be neglected \cite{StoeckmannBuch}.

As pseudointegrable systems obey the same kind of quadratic proliferation rule as integrable systems (cf. Eq.~(\ref{proliferat})), one can assume
that the diagonal approximation should apply by the same reasons.
However, by calculating the related quantity 
\begin{eqnarray}
\label{formfak}
&&\Phi(\ell)=\nonumber\\
&&  \left<\frac 1 {16 \pi^3} \sum_{i,j} \frac {a_i a_j}{\sqrt{\ell_i\ell_j}} \cos\left(\sqrt{E_0}(\ell_i-\ell_j)\right) \delta\left(\frac {2\ell-\ell_i+\ell_j}2\right) \right>, \nonumber \\
\end{eqnarray}
doubts on the validity of the diagonal approximation have been expressed \cite{biswas2}, because due to the different proportionality factors, $N(\ell)$ nevertheless increases faster than in integrable systems. 
Deviations between the diagonal approximation and the full summation have been found in \cite{biswas2} by investigating numerically the integral
\begin{eqnarray}\label{Ioftau}
&&I(\ell)= \frac{1}{2\sqrt{E_0}\lan d\ran^2}\int_0^{\ell_{\mathrm{max}}} \Phi(\ell')d\ell' \nonumber\\
&=& \left<\frac {1}{32\pi^3 \sqrt{E_0}  \left<d\right>^2} 
\sum_{i,j}^{(\ell_i + \ell_j)/2 < \ell} \frac {a_ia_j}{\sqrt{\ell_i\ell_j}}\cos \left(\sqrt{E_0} (\ell_i-\ell_j)\right)\right>, \nonumber \\
\end{eqnarray}
where $\ell_{\mathrm{max}} =\ell/(4\pi\sqrt{E_0}\lan d\ran)$ and the energy average runs again over several values of $E_0$. 

\unitlength 1.85mm
\vspace*{0mm}
\begin{figure}
\begin{picture}(0,30)
\def\epsfsize#1#2{0.65#1}
\put(0,0){\epsfbox{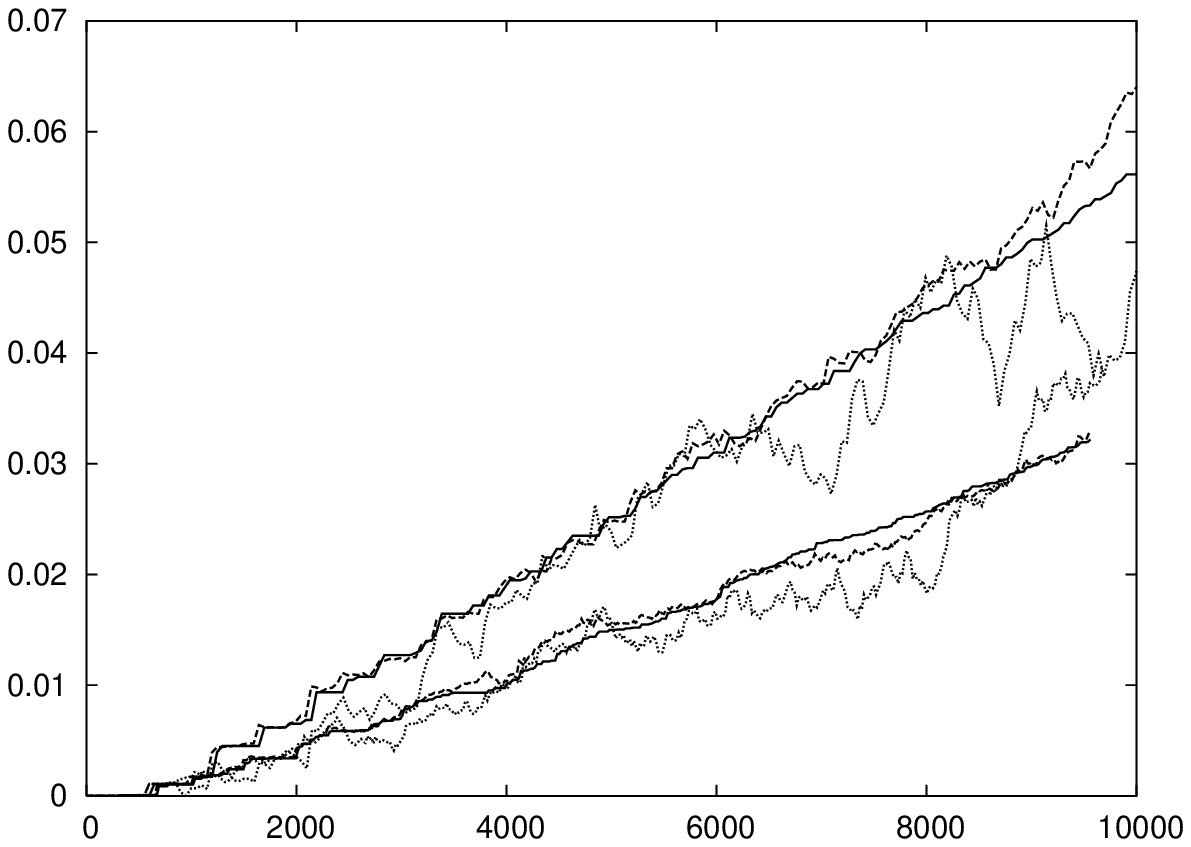}}
\put(23.5,-1.0){\makebox(1,1){$\ell$}}
\put(-1.0,25){\makebox(1,1){$I(\ell)$}}
\end{picture}
\caption{\small $I(\ell)$ for a rectangle (upper three curves) and a two-step system (lower three curves) (for details of the systems see Tab.~\ref{tab:1}). In both cases, the diagonal approximation $I(\ell)$ is close to a straight line (solid line), whereas the full sum oscillates around it (dotted and dashed lines). The oscillations are significantly reduced by averaging over more energy values. The sums are carried out over the first $3000$ orbits, including repetitions.}
\label{bi:itau}
\end{figure}

Here, we want to take a closer look to these deviations and therefore investigated $I(\ell)$ for our systems by calculating expression (\ref{Ioftau}) first in diagonal approximation and second by carrying out the full sum. 
When some orbit lenghts are degenerated, we include {\it all} pairs of orbits with $\ell_i=\ell_j$ into the diagonal approximation. This is equivalent to combining orbits with $\ell_i=\ell_j$ to one single orbit family with area $a_i+a_j$. 

The results are shown in Fig.~\ref{bi:itau}. The upper three curves show the  rectangular system and the lower ones the two-step system. Both systems are evaluated (i) in diagonal approximation (solid lines) and with the full sum and an average over (ii) $160$ (dashed lines) and (iii) $20$ values of $E_0$ (dotted lines), showing small and large fluctuations around the solid line, respectively. 
The results for the one-step system, that are qualitatively the same, are omitted for a better overview of the curves.

Contrary to \cite{biswas2}, we do not find a crossover value for $\ell$ in pseudointegrable systems above which the diagonal approximation breaks down. Instead we find fluctuations for $I(\ell)$ around the diagonal approximation $I_D(\ell)$ that depend on the energy average and increase with $\ell$. They can become quite large and seemingly distant from $I_D(\ell)$. 
However, these fluctuations occur in both, the integrable and the pseudointegrable systems and are significantly reduced by the average procedure. Therefore, the diagonal approximation seems to be valid for both kinds of systems, as long as the number of different $E_0$ values as well as their interval $\Delta E$ is not too small.   

\section{Spectral rigidity}
\label{sec_Delta3}
We finally investigate the spectral rigidity $\Delta_3(L)$. For integrable systems, it is known that $\Delta_3(L)$ shows linear behavior for small $L$ (see section I) and reaches a plateau for large $L$, the height of the plateau being determined by the smallest periodic orbit of the system. We now use the periodic orbit theory to find $\Delta_3(L)$ for our pseudointegrable systems.

One may ask, if the grouping of $\lan a \ran$ and $N(\ell)$, observed in section II also leads to an arrangement of the $\Delta_3$-curves according to their genus numbers. This can be roughly estimated from Eq.~(\ref{biswaseq}) by again investigating the neutral orbits. By going from the rectangle to the one-step system, the neutral orbit families splits into two new ones. With a characteristic step size of half the system size, one of the two new orbits occupies half the area $a_1$ than before while the area $a_2$ occupied by the other one is even smaller. Considering the length of the new orbits as roughly constant (which is of course only true for the first one), their common diagonal  contribution to $\Delta_3$ is $\propto (a_1^2+a_2^2)$, roughly half the contribution $\propto (2a_1)^2$ of the respective rectangular orbit. A similar discussion applies also for the more complicated orbits and for higher genus numbers. Therefore, $\Delta_3$ of a rectangle should be roughly diminished by a factor of two, when introducing a step of half the system size ($g=2$), and by a factor of three, when introducing  two steps of equal size ($g=3$).

However, the situation is very different when the step size is small compared to the system size. In this case, the contribution $a_1$ of the first new orbit is very close to the contribution of the rectangular orbit, whereas the contribution $a_2$ becomes negligible. Therefore, for small step sizes, the splitting of one orbit family into two new ones can be neglected and we expect $\Delta_3$ curves very close to the one of the rectangular system. In this case,  the $\Delta_3$ curves should not be ordered according to their genus numbers.

We now want to investigate these assumptions by our calculations.
First, $\Delta_3(L)$ is calculated by the diagonal approximation (open symbols) and by the full sum (solid symbols) and is plotted in Fig.~\ref{bi:Delta3} versus $L$ for the rectangular system (squares) and for our systems with $g=2$ (circles) and $g=3$ (triangles). The energy average was carried out over $160$ values of $E_0$ and the number of orbits, taken into account, is fixed to 3000 including repetitions. We carefully checked that the inclusion of even more orbits does not change the results of any of the considered systems significantly. The results can be summarized as follows: 

The calculated $\Delta_3(L)$ values for the rectangle are in quite good agreement with the theoretical slope of $L/15$ for small $L$. Also in the pseudointegrable cases, $\Delta_3(L)$ seems to increase linearly for small $L$ and saturates for larger $L$, but the slopes are smaller and decrease with increasing genus numbers, thereby coming closer to the $\Delta_3$ curve of the chaotic systems (lowest curve). 
The agreement between the diagonal approximation and the full sum is excellent. 

\unitlength 1.85mm
\vspace*{0mm}
\begin{figure}
\begin{picture}(0,30)
\def\epsfsize#1#2{0.65#1}
\put(0,0){\epsfbox{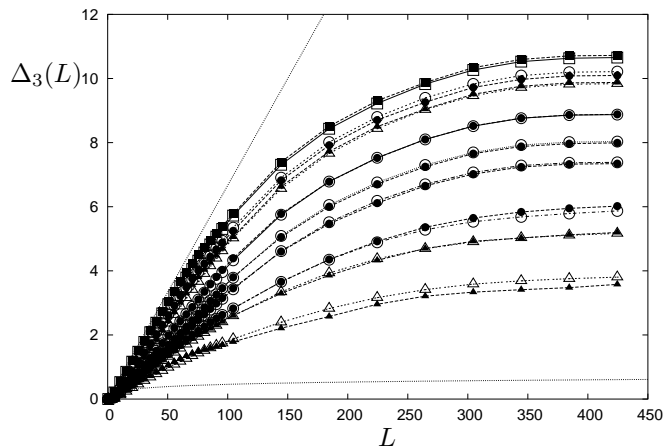}}
\put(23.5,-1.0){\makebox(1,1){$L$}}
\put(-1.0,25){\makebox(1,1){$\Delta_3(L)$}}
\end{picture}
\caption{\small $\Delta_3(L)$ is calculated by the diagonal approximation (open symbols) and by the full sum (solid symbols) and plotted versus $L$ for different systems of genus number $g=1$ (squares), $g=2$ (circles) and $g=3$ (triangles) (for details see Tab.~\ref{tab:1}).
The theoretical behavior for integrable and chaotic systems is indicated by the upper and the lower line, respectively (dotted lines without symbols).} 
\label{bi:Delta3}
\end{figure}

Next, we investigate the dependence on $g$.
When the step sizes are roughly equal in size, the curves of Fig.~\ref{bi:Delta3} show a systematic behavior on $g$, with $\Delta_3(L)$ decaying towards smaller values with increasing $g$. The curves for the rectangle are highest and most curves for the $g=2$ systems are above those with $g=3$. 
As expected from the discussion above, a systematic dependance on $g$ occurs also for the lowest curves of each genus number (refering to steps that split the system into parts of roughly equal size). These curves lie roughly by a factor of $2$ (for $g=2$) and of $3$ (for $g=3$) below the curves of the rectangle, which could be surprisingly well estimated by the behavior of the neutral orbits (see above).
However, for very small step sizes, this behavior changes and only slight deviations of $\Delta_3$ from the curve of the rectangle occur. This can be seen on the upper three curves, which refer to $g=1,2$ and $3$, but to very small step sizes and means that for given $g$ all curves between the $L/15$ line and a minimum curve can be found by changing the step sizes appropriately (the slope of the minimum curve depending on $g$). Therefore, beside the genus number $g$, the details of the specific system play an important role as well. 

It is however interesting to note that with fixed $g$ and increasing step sizes, the curves decay quite rapidly towards their minimum curve and that for small steps sizes, the changes in size have the strongest effect. For the top four curves of the one-step systems (circles), for example, the step size was varied over $10\%$ of the system size. The decrease in the slope is over $30\%$, and more then half of the maximum decrease. The very large change in step size from the 4th to the 5th curve, on the other hand, has a comparatively small effect. 
This is the reason that most of the curves shown in Fig.~\ref{bi:Delta3} show indeed a systematic dependance on $g$.

Another interesting quantity is the plateau value:
For large $L$, the values of $\Delta_3$ saturate to a plateau, whose height is determined by the small orbits. In order to estimate this height, one normally uses the approximation \cite{berry1} 
\beq\label{berryapprox}
\Delta_3(L)=2\lan d\ran\hbar\int_0^\infty\frac{dt}{t^2}\,\Phi(t)G(\frac{Lt}{2\lan d\ran\hbar})
\eeq
with $\Phi$ from Eq.~(\ref{formfak}), $t=\ell/(2E_0)$ and the so-called orbit selection function $G(x)$. $G(x)$ can roughly be approximated by a step function which is equal to $1$ in the plateau regime.
For integrable systems, $\phi(\ell)=\lan d\ran/(2\pi)$ for $\ell>\ell_{\rm{min}}$ (where $\ell_{\rm{min}}$ is the length of the shortest orbit of the system) and $\phi(\ell)=0$ for $\ell<\ell_{\rm{min}}$, which makes the calculation of $\Delta_3$ relatively easy \cite{berry1}. In the plateau regime, the integral (\ref{berryapprox}) depends only on the lower integration limit that is determined by the length of the shortest orbit of the system.

For our pseudointegrable systems, things become more complicated. We can find $\phi(\ell)$ from the slope of $I(\ell)$ in Fig.~\ref{bi:itau} in diagonal approximation, but the values differ for the different systems. This means that the behavior of the plateau depends on the length of the shortest orbit as well as on the system dependent value of $\Phi(\ell)$. 
As both quantities can be varied independentely from each other, there is no easy expression for the plateau value as in the case of the integrable systems.

Finally, in Fig.~\ref{bi:Delta3_ew}, we compare $\Delta_3^{\mathrm{PO}}(L)$ from the periodic orbit results (open symbols, diagonal approximation) to $\Delta_3^{\mathrm{e,L}}(L)$ from the eigenvalue statistics (filled symbols). To this end, we have calculated the eigenvalues under Neumann boundary conditions by the Lanczos algorithm (which involves a discretization of the lattice) and obtained $\Delta_3^{\mathrm{e,L}}(L)$ by using the technique as derived in \cite{bohigianno1975}. Our results are in line with previous numerical 
calculations on the two-step system, where the eigenvalues have been 
obtained for small values of $L$ by numerical diagonalization algorithms 
(\cite{cc}, \cite{biswas2} and \cite{biswas3}) and by the boundary element method (\cite{shudoshim} and \cite{shudoetal}). For the rectangle, also the exact eigenvalues for a continuous system are used for comparision. 
For the rectangle, the $\Delta^{\mathrm{e}}_3$ curve as obtained from the exact eigenvalues (solid diamonds) 
agrees very well with $\Delta^{\mathrm{PO}}_3$ from the periodic orbit theory. 
This had to be expected for a system without diffractive corners. 
The $\Delta_3^{\mathrm{e,L}}$ curve, obtained from the numerical eigenvalues of a discretized systems lies somewhat higher
and the deviation gives us the error bar arising from the discretization.

The results of several pseudointegrable systems of $g=2$ and $g=3$ systems are shown in Fig.~\ref{bi:Delta3_ew} as well. Also in these cases, the agreement between $\Delta^{\mathrm{e,L}}_3$ and $\Delta^{\mathrm{PO}}_3$ is quite good.
Again, $\Delta^{\mathrm{e,L}}_3$ lies slightly above $\Delta_3^{\mathrm{PO}}$, which can in principle be due either to the discretization or to the neglection of higher-order terms, as e.g. the diffraction terms. The deviation seems even smaller than for the rectangle. However, by calculating rectangles of different sizes, we found that the deviations between $\Delta^{\mathrm{e,L}}_3$ and $\Delta_3^{\mathrm{PO}}$ are in most cases smaller than the one shown here, which we therefore consider as the upper limit of the error due to discretization. 

\unitlength 1.85mm
\vspace*{0mm}
\begin{figure}
\begin{picture}(0,30)
\def\epsfsize#1#2{0.65#1}
\put(0,0){\epsfbox{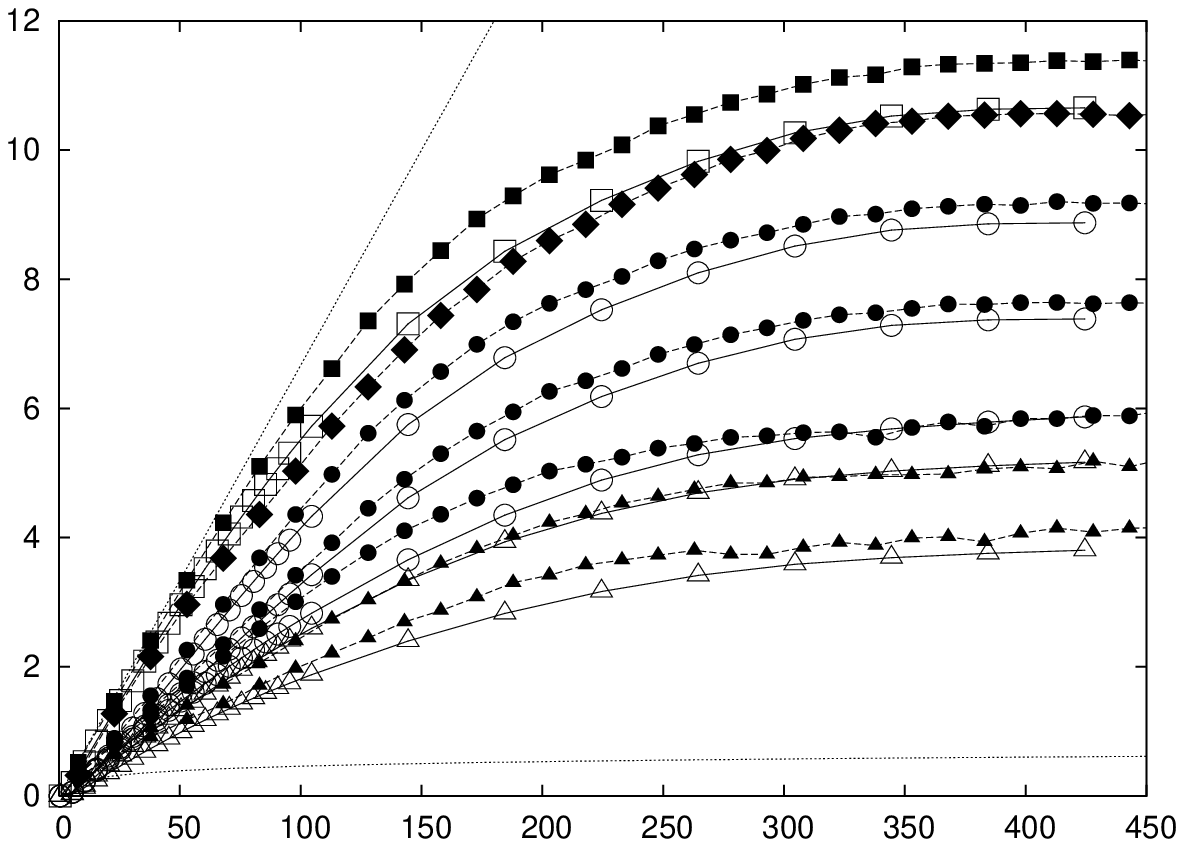}}
\put(23.5,-1.0){\makebox(1,1){$L$}}
\put(-1.0,25){\makebox(1,1){$\Delta_3(L)$}}
\end{picture}
\caption{\small $\Delta^{\mathrm{PO}}_3(L)$ from the periodic orbit theory in diagonal approximation (open symbols) is compared to $\Delta^{\mathrm{e,L}}_3$ calculated from the  eigenvalues under Neumann boundary conditions obtained by the Lanczos algorithm (solid symbols) for different systems.  For the rectangle, $\Delta^{\mathrm{e}}_3$ as calculated from the exact eigenvalues for a continous rectangluar system is shown as well (solid diamonts). Different symbols represent the genus numbers $g=1$ (squares), $g=2$ (circles) and $g=3$ (triangles).
The theoretical behavior for integrable and chaotic systems is indicated by the upper and the lower line, respectively (dotted lines without symbols).
}
\label{bi:Delta3_ew}
\end{figure}

On the basis of the present data, it seems that the contribution of the higher-order terms is quite small and at most in the same order of magnitude as the errors due to discretization. It is also interesting to note that the deviations do not increase, when going from $g=2$ to $g=3$, which doubles the number of diffractive corners. This may also be a hint that the diffraction terms are not very important in our systems,
where the diffraction terms are arising from the corner scattering. The situation seems to be different in systems with point-like scatterers, where the spectral statistic is known to be changed drastically\cite{seb90}.
It will be interesting to investigate systems with higher $g$ to see, if this assumption holds. 

\section{Summary and Conclusions}\label{sec_Concl}

In summary, we have calculated the spectral rigidity $\Delta_3(L)$ for various pseudointegrable systems by applying periodic orbit theory. By averaging over enough energy values taken from a not too small interval, we found that the diagonal approximation is applicable. 
Most $\Delta_3(L)$ curves decrease with increasing genus number $g$, but other details of the geometry play an important role as well. In particular, when the salient corners are very small, all curves approach the one of the rectangle, independentely of their genus numbers.

The behavior of the different $\Delta_3(L)$ curves can be understood in terms of the periodic orbit families. Estimating the behavior of the orbits from the neutral orbits that bounce between two parallel walls gives already a crude approximation of the lowest curves of a given $g$ that works surprisingly well. This estimation also shows that the lowest curves occur for systems with salient corners of roughly equal size. If the corners of the systems are very small, on the other hand, the opposite behavior occurs and the $\Delta_3(L)$ curves lie very close to the one for a rectangle. Also this can be quite well understood by looking at the periodic orbit families. The ones that possess the same lengths as the rectangle occupy in this case the dominant areas in phase space, while the orbit families that differ from the ones of the rectangle become negligible.

Finally, we discuss which effects have been neglected and could lead to deviations from our results. 
First, there are various attempts in literature to go beyond periodic orbit theory, 
especially the attempt to take the influence of diffractive orbits into account. These orbits start at the singularities, in our case at the salient corners, and their contributions to spectral statistics have been treated in \cite{sieber99}. They are in general non-negligible (except for $L\to 0$), but smaller than the periodic orbit contributions. 
Second, there might be an effect of the larger orbits beyond the first 3000 ones. However, as we discussed in section III, their influence should be negligible. Indeed, the difference between $\Delta_3$ calculated by taking 3000 or 12000 orbits (including repetitions) was hardly visible. 

We can get hints about the importance of the diffractive orbits by comparing the $\Delta_3$ curves obtained from the eigenvalues and the ones obtained from the periodic orbit calculations. As shown in the last section, the agreement between both is quite good. Moreover, the difference between the periodic orbit results and the eigenvalue results is roughly the same in all considered systems, independentely of number and size of the salient corners. This indicates that the (small) difference between both methods is rather due to the discretization of the eigenvalue calculations. Therefore, in agreement with \cite{biswas2} we think that the influence of diffractive orbits to spectral statistics in our systems is only small.
Of course, it is possible that it increases when still more corners 
and/or additional diffraction terms, arising e.g. from flux lines or point like scatterers, 
are present.

Moreover, it will also  be very interesting to investigate the connection between the $\Delta_3$ curves and the periodic orbits for more complicated systems with higher genus numbers. Introducing more and more corners leads to more and more different orbit families with quite similar lengths, but rather small areas in phase space. When many -- even small -- corners are present, it is quite obvious that the original orbits of the rectangle are considerably disturbed, i.e. that their area in phase space becomes much smaller. Therefore, contrary to the relatively simple systems of this work, one should expect a considerable change of $\Delta_3$. Many numerical works have shown that the curves are shifted towards the one of chaotic systems and it will be very interesting to rely this shifting also to the periodic orbit families of the respective systems.

\section{Acknowledgments}
\label{sec_Acknow}
We gratefully acknowledge financial support from the Deutsche 
Forschungsgemeinschaft and valuable discussions with Yuriy Hlushchuk.

\end{multicols}

\begin{table}
\begin{tabular}{|l|c|c|c|c|c|c|c|c|c|c|c|c|} \hline
      & \multicolumn{6}{c|}{\rule[-2mm]{0mm}{6mm}System Lengths}
      & \multicolumn{6}{c|}{\rule[-2mm]{0mm}{6mm}Figure} \\ \hline 
      & {\rule[-2mm]{0mm}{6mm}} $X_1$& $X_2$ &$X_3$ &$Y_1$& $Y_2$ &$Y_3$ &\ref{bi:sumrule}   &\ref{bi:flaeche} & \ref{bi:proliferat} & \ref{bi:itau}  & \ref{bi:Delta3} & \ref{bi:Delta3_ew} \\  \hline 
Rectangle & 307 & - & - & 503 & - & -                & 5  & 1 & 5 & 1-3& 1 & 1\\
One-step& 307 & 305 & - & 503 & 501 & -              &    &   &   &    & 2 & \\
One-step& 307 & 293 & - & 503 & 491 & -              &    &   &   &    & 4 & 2\\
One-step& 307 & 283 & - & 503 & 479 & -              & 4  & 3 & 4 &    & 5 & \\
One-step& 307 &  271 & - & 503 & 467 & -             &    &   &   &    & 6 & 3\\
One-step& 307 &199  & - & 503 & 397 & -              & 3  & 2 & 3 &    &   & \\
One-step& 307 & 151 & - & 503 &347  & -              &    &   &   &    & 7 & 4 \\
Two-step& 307 & 305 & 303 & 503 & 501 & 499          & 2  & 4 & 2 &    & 3 &  \\ 
Two-step& 307 & 156 & 154 & 503 & 501 & 352          &    &   &   & 4-6& 8 & 5  \\
Two-step& 307 & 206 & 104 & 503 & 401 & 300          & 1  & 5 & 1 &    & 9 & 6  \\ \hline
\end{tabular}
\caption[]{\small Table of the systems used in this work. $X_i$ and $Y_i$ are the lengths of the different segments of the systems (also lengths of the ''neutral orbits'') as shown in Fig.~\ref{bi:geo}. The last columns indicate in which figures the different systems occur and the numbers given in these columns enumerate the curves from top to bottom.
\label{tab:1}}
\end{table}


\begin{thebibliography} {50}
\bibitem {footnote1} Eq.~(\ref{biswaseq}) differs from the one of Ref.~\cite{biswas2} by a factor of two, which is probably due to a different treatment of orbits in forward and backward direction.
\end{thebibliography}
\end{document}